\documentclass{emulateapj}

\shorttitle{SLACS VII}
\shortauthors{Bolton et al.\ }
 
\begin{document}

\title{The Sloan Lens ACS Survey. VII\@.
Elliptical Galaxy Scaling Laws from
Direct Observational Mass Measurements\altaffilmark{1}}

\author{Adam S. Bolton\altaffilmark{2,3}}
\author{Tommaso Treu\altaffilmark{4,5}}
\author{L\'{e}on V. E. Koopmans\altaffilmark{6}}
\author{Rapha\"el Gavazzi\altaffilmark{4,7}}
\author{Leonidas A. Moustakas\altaffilmark{8}}
\author{Scott Burles\altaffilmark{9}}
\author{David J. Schlegel\altaffilmark{10}}
\author{Randall Wayth\altaffilmark{3}}

\slugcomment{Accepted for publication in The Astrophysical Journal}

\altaffiltext{1}{Based on observations made with the
NASA/ESA 
\textsl{Hubble Space Telescope}, obtained
at the Space Telescope Science Institute, which is operated
by AURA, Inc., under NASA contract NAS 5-26555.
These observations are associated with programs \#10174, \#10494,
\#10587, \#10798, and \#10886.}
\altaffiltext{2}{Beatrice Watson Parrent Fellow,
Institute for Astronomy, University of Hawai`i,
2680 Woodlawn Dr., Honolulu, HI 96822, USA ({\tt bolton@ifa.hawaii.edu})}
\altaffiltext{3}{Harvard-Smithsonian Center for Astrophysics, 60 Garden
St., Cambridge, MA 02138, USA ({\tt rwayth@cfa.harvard.edu})}
\altaffiltext{4}{Department of Physics, University of California,
 Santa Barbara, CA 93101, USA ({\tt tt@physics.ucsb.edu})}
\altaffiltext{5}{Sloan Fellow, Packard Fellow}
\altaffiltext{6}{Kapteyn
 Astronomical Institute, University of Groningen, P.O. Box 800, 9700AV
 Groningen, The Netherlands ({\tt koopmans@astro.rug.nl})}
\altaffiltext{7}{Institut d'Astrophysique de Paris, UMR7095 CNRS \&
Univ.\ Pierre et Marie Curie, 98bis Bvd Arago, F-75014 Paris, France
(\texttt{gavazzi@iap.fr})}
\altaffiltext{8}{Jet Propulsion Laboratory, California Institute of
Technology, 4800 Oak Grove Drive, M/S 169-327, Pasadena, CA 91109, USA
({\tt leonidas@jpl.nasa.gov})}
\altaffiltext{9}{Department of Physics and Kavli Institute for
Astrophysics and Space Research, Massachusetts Institute of Technology,
77 Massachusetts Avenue, Cambridge, MA 02139, USA ({\tt burles@mit.edu})}
\altaffiltext{10}{Physics Division, Lawrence Berkeley National Laboratory,
Berkeley, CA 94720-8160, USA ({\tt djschlegel@lbl.gov})}

\begin{abstract}
We use a sample of 53 massive early-type strong gravitational lens galaxies
with well-measured redshifts (ranging from $z=0.06$ to 0.36)
and stellar velocity dispersions (between 175 and 400\,km\,s$^{-1}$)
from the Sloan Lens ACS (SLACS) Survey to derive numerous empirical
scaling relations.
The ratio between central stellar velocity
dispersion and isothermal lens-model velocity
dispersion is nearly unity within errors.
The SLACS lenses define a
fundamental plane (FP) that is consistent with the FP of
the general population of early-type galaxies.
We measure the relationship between
strong-lensing mass $M_{\mathrm{lens}}$ within one-half
effective radius ($R_e/2$) and the dimensional mass variable
$M_{\mathrm{dim}} \equiv G^{-1} \sigma_{e2}^2 (R_e / 2)$ to be
$\log_{10} [M_{\mathrm{lens}}/10^{11} M_{\odot}]
= (1.03 \pm 0.04) \log_{10} [M_{\mathrm{dim}}/10^{11} M_{\odot}]
+ (0.54 \pm 0.02)$ (where $\sigma_{e2}$
is the projected stellar velocity
dispersion within $R_e/2$).  The near-unity slope
indicates that the mass-dynamical structure of massive
elliptical galaxies is independent of mass, and that the
``tilt'' of the SLACS FP is due entirely to variation
in total (luminous plus dark) mass-to-light ratio with mass.
Our results imply that dynamical masses serve as a good proxies
for true masses in massive elliptical galaxies.
Regarding the SLACS lenses as a homologous population,
we find that the average enclosed 2D mass profile
goes as $\log_{10} [M(<\!\!R)/M_{\mathrm{dim}}] =
(1.10 \pm 0.09) \log_{10} [R/R_e] + (0.85 \pm 0.03)$,
consistent with an isothermal (flat rotation curve) model
when de-projected into 3D\@.
This measurement is inconsistent with the slope of the average projected
aperture luminosity profile at a confidence level greater than 99.9\%,
implying a minimum dark-matter fraction of $f_{\mathrm{DM}} = 0.38 \pm 0.07$
within one effective radius.  We also present an analysis
of the angular mass structure of the lens galaxies,
which further supports the need for dark matter inside one
effective radius.
\end{abstract}


\section{Introduction}

Elliptical galaxies are simple in appearance,
but their internal structure is
resistant to elementary Newtonian
deduction because the primary luminous
tracers of their gravity---stars---move not on ``cold''
circular orbits but on ``hot'' randomized orbits with a broad
distribution in phase space
\citep[e.g.,][]{bertola_capaccioli_75, binney_76, binney_78, illingworth_77}.
With detailed spatially resolved
observations of the absorption-line kinematics of nearby elliptical
galaxies, this difficulty can be attacked head-on through the
use of dynamical modeling of the distribution function of stars
in phase space (e.g., \citealt{vdm_91}; \citealt{rix_white_92};
\citealt*{bss_92a}; \citealt*{bss_92b};
\citealt{saglia_93}; \citealt{vdm_franx_93};
\citealt{merritt_saha_93}; \citealt{kuijken_merrifield_93}; \citealt{gerhard_93};
\citealt{bsg_94}; \citealt{bertin_94}; \citealt{rix_97};
\citealt{kronawitter_2000}; \citealt{romanowsky_m87};
\citealt{saglia_2000}; \citealt{kronawitter_2000};
\citealt{gerhard_2001} \citealt{capp_sauron}).  Such efforts require a very detailed
level of analysis, and rely on assumptions of dynamical
relaxation and the choice of correct dynamical model families.

Strong gravitational lensing offers an attractive complement
to dynamical modeling, since it measures the total mass
within the ``Einstein radius'' defined by the lensed
images---a scale that is generally comparable to the effective
radius of the luminous matter distribution in the case of galaxy-scale lenses.
These lensing mass measurements are extremely robust, with few underlying
assumptions and minimal model dependence \citep[e.g.,][]{kochanek_91}.
The main limitation to
strong lensing as a probe of elliptical galaxy structure has typically
been the lack of sufficiently large and uniform lens samples.
This limitation has been overcome in recent years by the Sloan Lens ACS
Survey (SLACS: \citealt{slacs1, slacs2, slacs3, slacs4, slacs5, slacs6};
hereafter Papers~I--VI) and other surveys
\citep[e.g.,][]{winn_pmn_i, winn_pmn_ii, winn_pmn_iii, winn_pmn_iv,
myers_class, browne_class, maoz_snap, morgan_ctq327, he_lens_i, he_lens_ii,
he_lens_iii, he_lens_iv, he_lens_v, he_lens_vi, he_lens_vii, ols1, ols2,
sdssqls1, sdssqls2, sdssqls3, mds_lens, lam_legs,
cabanac_sl2s, belokurov_lens, kubo_lens, faure_cosmos}, which
have begun to deliver statistically significant samples of strong
lens galaxies by using systematic search techniques.

Previous studies have taken diverse approaches to the use
of strong-lensing information for the measurement
of early-type galaxy mass-density structure.
Strong-lensing aperture
masses have been used by \citet*{rkk_03} and \citet{rusin_kochanek_05} to provide
ensemble constraints upon the mass-density structure
of early-type galaxies.  Lensing mass constraints have been
combined in self-consistent fashion with stellar dynamics
to constrain the mass structure of lenses on a system-by-system
basis by the Lenses Structure
and Dynamics Survey (LSD: \citealt{kt02, kt03, tk02, tk03, tk04})
and the SLACS Survey (Paper~III).  \citet{kochanek_what} and \citet{kochanek_morgan_06}
have used the time delays between multiple lensed quasar images, together
with a given value for the Hubble parameter $H_0$ to measure the local
mass slope of the lensing galaxies.  Constraints on the mass profile
of lenses based upon resolved lensed features have been published
by \citet{cohn_2001}, \citet{munoz_2001},
\citet{wayth_0047}, \citet{dye_warren_05}, \citet{brewer_06},
and \citet{dye_lbg}.  \citet[][hereafter B07]{bolton_mfp}
have used lensing aperture masses in combination with central stellar
velocity dispersions and measured effective radii to demonstrate
on an empirical scaling basis
that the mass-density structure of massive elliptical lenses
from the SLACS Survey is on average independent of mass,
a result which we establish and explore in more detail in the
current work.

In this work, we employ the mass, light, and velocity measurements
of the full
\textsl{Hubble Space Telescope} (\textsl{HST}) Advanced Camera
for Surveys (ACS)
gravitational lens sample from the SLACS Survey, published
in Paper~V, to derive numerous empirical scaling relations.
The analysis of the initial SLACS sample in Paper~III
allowed us to measure an on-average isothermal mass-density structure;
the current lens sample is large enough to investigate \textit{trends} in
structure with mass and velocity dispersion.
We frame much of our analysis and discussion in terms
of the Fundamental Plane scaling relation
(FP; e.g., \citealt{dd_fp, dr_fp, bbf_92, renzini_ciotti_93,
guzman_93, pahre_95, pahre_iii, pahre_iv, jorgensen_fp,
vd_franx_96, clr_fp_96, kelson_97, graham_colless_97,
scodeggio_98, kochanek_lens_fp, treu_2001_iii, treu_2002, bcdp_02,
vd_stanford, bernardi_fp, vandeven_fp, tbb_fp, vanderwel_05}; Paper~II; B07)
and the relationships between luminosity,
``dynamical mass'', and strong-lensing aperture mass
that illuminate the structural explanation for the
``tilt'' of the FP\@.
In the SLACS sample, we have a large number of early-type galaxies
distributed across the higher-mass
end of the FP, with uniform and high-quality
measurements of the observables of redshift,
surface brightness, velocity dispersion, and effective radius.
\textit{In addition to these quantities, we have a full set
of aperture masses---measurements of the total mass within the
Einstein ring radius.}  These additional mass measurements
add another dimension of physical constraint to the SLACS
sample that is not available for other FP galaxy samples,
thus offering the opportunity to break some of the degeneracy
inherent in the physical interpretation of FP analyses.

This paper is arranged as follows.  Section~\ref{overview} gives an
overview of our measurements and analysis techniques.
Sections~\ref{angular}--\ref{mml} present the various scaling relations
defined by the SLACS sample, including the relative alignment and flattening
of projected mass and light distributions (\S\ref{angular}),
the relation between stellar and lensing velocity-dispersion
measurements (\S\ref{v_v}),
the FP and ``mass plane'' relations (\S\ref{fpmpsec}), and various
perspectives on the relationship between dynamical mass, lensing mass, and luminosity
(\S\ref{mml}).  In \S\ref{mofr} we make a robust determination
of the ensemble average radial enclosed-mass profile of the SLACS sample.
Section~\ref{summary} provides an itemized summary of our results,
and \S\ref{discuss} provides a concluding discussion.

All computations in this work assume a general relativistic
Friedmann-Robertson-Walker cosmology with matter-density
parameter $\Omega_{\mathrm{M}} = 0.3$, vacuum energy-density
parameter $\Omega_{\Lambda} = 0.7$, and Hubble parameter
$H_0 = 70$\,km\,s$^{-1}$\,Mpc$^{-1}$.

\section{Overview of sample, measurements, and methods}
\label{overview}

We consider the sample of 63 ``grade-A''
strong gravitational lenses presented in Paper~V\@.  For analyses
employing stellar velocity dispersions, we
restrict the sample to the 53 early-type
lens systems with a median SDSS spectral
signal-to-noise ratio of 10 or greater per 69\,km\,s$^{-1}$ pixel
over the rest-frame range of 4100 to 6800\AA\@.
The measurements upon which the current work is based
are all presented in Paper~V, and are described
only briefly here.

We use stellar velocity dispersions measured from Sloan Digital Sky
Survey (SDSS: \citealt{york_sdss}; \citealt{dr6}) spectroscopy,
which samples a 3$\arcsec$-diameter circular fiber aperture centered on
the target galaxy.  We correct these measured
velocity dispersions (which we denote by $\sigma_{\mathrm{SDSS}}$)
using the empirical power-law relation of \citet{jorgensen_vdisp}
to uniform physical apertures of either $R_e / 8$
(to give a ``central'' velocity dispersion $\sigma_{e8}$) or $R_e / 2$
($\sigma_{e2}$, for a spatial aperture more closely matched to that
of the spectroscopy and of the strong-lensing features).
We note that these corrections
are quite small---RMS of 2.5\% for $R_e / 2$---and that
the coefficients of logarithmic scaling relations
we derive are not sensitive to the particular choice
of velocity-normalization aperture.
We use magnitudes,
effective radii, projected axis ratios, and
position angles for the lens galaxies
measured from fitting de Vaucouleurs
models to ACS-WFC F814W imaging.
To convert from observed magnitudes to
rest-frame luminosities, we first correct for
Galactic dust extinction using the maps of
\citet*{sfd_dust}, and apply $k$-corrections and evolutionary
corrections.
Our strong gravitational-lens models provide for each lens a
robust measurement of the mass $M_{\mathrm{Ein}}$ enclosed within
the physical Einstein radius $R_{\mathrm{Ein}}$ (derived from the angular
Einstein radius $b$ together with redshifts and cosmology),
for both singular isothermal ellipsoid
(SIE: \citealt{kassiola_kovner}; \citealt{kormann_sie};
\citealt{keeton_kochanek})
and light-traces-mass (LTM) models for the
lens-galaxy mass-density profile.
For much of our analysis,
we evaluate the strong-lensing aperture masses for a uniform physical
aperture of $R_e / 2$ (one-half the effective radius of the
luminous component), chosen to be
fairly closely matched to the sample median
Einstein radius $R_{\mathrm{Ein}}$ of
approximately $0.6 R_e$.  The evaluation of the
aperture mass at a radius other than $R_{\mathrm{Ein}}$
introduces some dependence upon the assumed lens-mass model;
we gauge this effect by
using aperture masses from both
SIE and LTM models, by checking for systematic
residual correlations with the ratio $R_{\mathrm{Ein}} / R_e$
(of which we find none), and by verifying that our results
do not change when derived using only the half of the sample with the
smallest fractional aperture-mass difference between
$R_{\mathrm{Ein}}$ and $R_e / 2$.
For an overview of the theory, phenomenology, and scientific applications
of strong gravitational lensing, we refer the reader to
Part~2 of \citet*{saas_fee}.

Unless otherwise noted, all scaling relations
are fitted as linear and plane
relationships in logarithmic space.  We define the best fit as
that which minimizes the total squared orthogonal distance from
the line or plane to the set of sample data points.
Before carrying out the fits, we scale
all logarithmic data coordinates
by their typical logarithmic errors,
so as to apply a
roughly uniform error metric across the multiple observables
that span the spaces under consideration.
The adopted errors are taken from the comparisons among multiple
measurement techniques in Paper~V, and are given in Table~\ref{dataerror}.
We estimate errors
on our fitted parameters using bootstrap re-sampling of the
analysis sample \citep{efron_bootstrap}.
Our reported errors
are the square-root diagonal entries
of the parameter covariance matrices constructed from the
sets of bootstrap parameter estimates.

\begin{table}[t]
\begin{center}
\caption{\label{dataerror} Adopted errors in measured data values}
\begin{tabular}{lc}
\hline \hline
Measured parameter & Adopted error \\
\hline
$\log_{10}$ Luminosity $L_V$ & 0.010\,dex \\
$\log_{10}$ Effective radius $R_e$ & 0.015\,dex \\
$\log_{10}$ Aperture mass $M$ & 0.010\,dex \\
$\log_{10}$ Surface brightness $I_e=L_V/(2 \pi R_e^2)$ & 0.020\,dex \\
$\log_{10}$ Velocity dispersion $\sigma$ & 0.030\,dex \\
\hline
\end{tabular}
\end{center}
\end{table}

\section{The comparative angular structure of mass and light}
\label{angular}

In addition to the measurement of aperture masses,
the simplest strong-lens models also give a determination
of the angular structure of the lens mass profile, in the form of
either a mass axis ratio (for the SIE models) or an external
shear magnitude (for the LTM models), along with the
associated position angle.
The implications of these measurements were discussed at length
in Paper~III; we briefly revisit the analysis here for the
larger sample of lenses presented in Paper~V\@.

First we examine the position-angle alignment between the major axes
of the light profile (as determined by the de Vaucouleurs fit) and the mass
profile (as determined from the SIE model). %,
For the 58 lenses with
light-profile axis ratios $q_{\mathrm{stars}} < 0.95$,
the mean position-angle difference
between mass and light profiles is $2.5^{\circ} \pm 2.4^{\circ}$.
The RMS alignment is $18^{\circ}$, as compared with an intrinsic
error of $7.5^{\circ}$ in the mass position angles from the analysis of Paper~V\@.
We can further restrict our attention to a subset
of 29 of these lenses whose lensed images extend over a significant azimuth about
the lens-galaxy center, for which the mass PA measurement
is more tightly constrained, taken from the 32 lenses
of the ``ring subset'' described in Paper~V\@.  In this case,
we find a mean mass-light PA difference of $1.4^{\circ} \pm 1.9^{\circ}$.
From Paper~V, the RMS error in the mass position angle measurement
for the ring subset is approximately $2^{\circ}$.
The RMS position-angle difference for the ring subset of lenses
is $10^{\circ}$, equal to the value found in Paper~III for the original SLACS sample.

Interestingly,
the LTM plus external shear models of Paper~V
show a significant amount of preferential alignment of
the ``external'' shear with the
major-axis PA of the de Vaucouleurs surface-brightness
model.  Of the above-mentioned 29 well-constrained (``ring subset'') lenses
with $q_{\mathrm{stars}} < 0.95$, 17 (59\%) have their
shear PA aligned to within $\pm 15^{\circ}$ of their luminosity PA, an
angular range that only encompasses 17\% of the meaningful range
of variation ($\pm 90^{\circ}$).  If the shear were external and uncorrelated
with the PA of the luminous distribution, this level of random alignment
would have a probability of less than one in $10^6$.  This result suggests
that much of the shear required by the LTM lens models is not of
environmental origin, but is instead compensation for an intrinsic shortcoming
of the LTM lens models.  The fact that the shear is aligned with the light
position angles rather than being anti-aligned ($90^{\circ}$ out
of phase) shows that the lensed features require a stronger gravitational
quadrupole at the locations of the lensed images than can be provided by
the pure LTM models.  This translates
into a requirement either that the radial mass-density profile be less centrally
concentrated than the light profile, or that the projected mass profile be
flatter than the projected luminosity profile.  In either case,
the improbable alignment of ``external shear'' with the major axes of LTM lens
models provides evidence against the light-traces-mass hypothesis and thus
in favor of some form of dark-matter component.

We also examine the relative flattening of mass and light
as measured from SIE and de Vaucouleurs models.
For the 57 grade-A lenses with early-type morphology, we find
$\left< q_{\mathrm{SIE}} / q_{\mathrm{stars}} \right> = 1.02 \pm 0.02$
with an RMS deviation of 0.12 about the mean: i.e., mass and light
have essentially the same projected axis ratio.
As in Paper~III, we see a trend from mass being rounder than light
at lower masses towards mass being flatter than light
at higher masses.
Considering the subset of grade-A lenses with early-type morphology
that both (1) are in the angularly well-constrained ``ring subset'',
and (2) have mass and light position angles aligned to within 15$^{\circ}$,
this trend can be seen as a linear correlation coefficient of
$q_{\mathrm{SIE}} / q_{\mathrm{stars}}$ with $\sigma_{\mathrm{SIE}}$ of
$r = -0.665$, corresponding to a significance of 99.95\%.

\section{Velocity dispersions: stellar and lensing}
\label{v_v}

For SIE lens models, we may translate the measured
Einstein radii $b$ into lens-model velocity dispersions
$\sigma_{\mathrm{SIE}}$ through the relation
\begin{equation}
\label{b_sie}
b_{\mathrm{SIE}} = 4 \pi {{\sigma_{\mathrm{SIE}}^2} \over {c^2}}
{{D_{\mathrm{LS}}} \over {D_{\mathrm{S}}}}~,
\end{equation}
where $D_{\mathrm{LS}}$ and $D_{\mathrm{S}}$ are ``angular-diameter distances''
from lens (foreground galaxy) to source (background galaxy)
and observer to source respectively.
Though the conversion is strictly appropriate only
for an isotropic, spherically symmetric, single-component density
distribution with $\rho \propto r^{-2}$ (in 3D),
it provides a robust
velocity-scale measurement with extremely small
statistical error, to be compared with the (usually noisier) central
velocity dispersion $\sigma_{\mathrm{stars}}$ of the stars within the lens galaxy.
By considering bulge-plus-isothermal-halo models,
\citet{kochanek_dyn_halo_94} suggested that an approximate
equality should hold between the central stellar line-of-sight
velocity dispersion and the velocity-dispersion parameter of the
isothermal halo.  This equality has generally been confirmed
in successively larger samples of lens galaxies with measured
stellar velocity dispersions (\citealt{kochanek_lens_fp}, LSD, Papers II and~III), and
the current SLACS lens sample allows the most significant determination to date.

\begin{figure}
\plotone{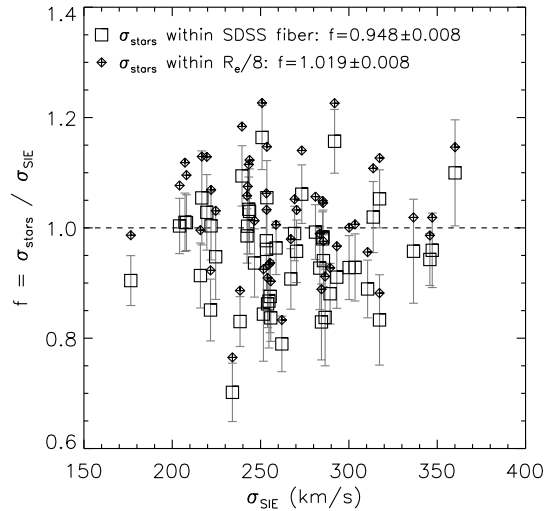}
\caption{\label{vvplot}
Ratio $f$ of stellar velocity dispersion $\sigma_{\mathrm{stars}}$
to velocity-dispersion parameter $\sigma_{\mathrm{SIE}}$ of the best-fit lens model,
as a function of $\sigma_{\mathrm{SIE}}$. % (left)
Points are shown for $\sigma_{\mathrm{stars}}$ as observed within the $3\arcsec$-diameter
SDSS fiber, and as corrected to a
uniform physical aperture of $R_e / 8$ using the
empirical formula of \citet{jorgensen_vdisp}.
For visual clarity, error bars (gray) are only
plotted for the fiber-aperture points.}
\end{figure}

Weighting with the statistical errors of the SDSS velocity dispersions
(limited to a minimum error-bar size of
$\delta \sigma_{\mathrm{stars}} = 0.05 \sigma_{\mathrm{stars}}$),
we find $\sigma_{\mathrm{SDSS}} = (0.948 \pm 0.008) \sigma_{\mathrm{SIE}}$ and
$\sigma_{e8} = (1.019 \pm 0.008) \sigma_{\mathrm{SIE}}$.
Thus we see that the scaling $\sigma_{e8} = \sigma_{\mathrm{SIE}}$
holds to within less than 3\%, and is essentially consistent with being exactly
true on average at the current level of evidence.  The RMS scatter
about the relation is approximately 0.09 for $\sigma_{\mathrm{SDSS}}$
and 0.1 for $\sigma_{e8}$, somewhat larger than seen in Paper~II\@.
The reduced $\chi^2$ of this scatter when weighted by the errors as in
the averaging is 2.7, showing evidence for intrinsic RMS scatter at
the level of about 20\,km\,s$^{-1}$, or alternatively at the
level of 7.5\% $\sigma_{\mathrm{SIE}}$ (i.e., the necessary value added
in quadrature to the velocity errors to give a reduced $\chi^2$ of
approximately unity).
Figure~\ref{vvplot} shows the ratio
$f \equiv \sigma_{\mathrm{stars}} / \sigma_{\mathrm{SIE}}$ as a function of
$\sigma_{\mathrm{SIE}}$.
Note that our definition of $f$ follows
the convention of Paper~II, but is the inverse of the definition used
in \citet{kochanek_lens_fp}.

The residuals about the mean
$\sigma_{\mathrm{stars}}$-versus-$\sigma_{\mathrm{SIE}}$
value are not significantly
correlated with $\sigma_{\mathrm{SIE}}$, mass, luminosity,
mass-to-light ratio, effective radius, or ratio
of Einstein radius to effective radius.
There is an \textit{apparent} correlation with $\sigma_{\mathrm{stars}}$,
but this is an artifact of the relatively large
statistical errors in that quantity.
The lack of a correlation between $f$
and lens mass in particular indicates that the
mass-dynamical structure of massive early-type galaxies is
independent of mass.  Likewise, the lack of correlation between
$f$ and the ratio of Einstein radius $R_{\mathrm{Ein}}$
to effective radius $R_e$ indicates that the
universal density structure is nearly isothermal
independently of dynamical modeling.  These
two results will be derived quantitatively
in \S\ref{mml} and \S\ref{mofr} below, and
discussed further in \S\ref{discuss}.

\citet{auger_07} has reported a significant
correlation for the 15 SLACS lenses analyzed in Paper~III between
the logarithmic mass-profile slopes (from a
combined lensing and dynamical analysis)
and the number of near neighbor galaxies to the lenses.
The (small) intrinsic scatter of the individual mass slopes about
the average value is directly related to the scatter in
the $\sigma_{e8}$--$\sigma_{\mathrm{SIE}}$ relation;
it remains to be determined whether the \citet{auger_07} result
is confirmed by the larger current SLACS sample (Treu et al., in preparation).

\section{The fundamental plane and mass plane}
\label{fpmpsec}

The fundamental plane (FP: \citealt{dd_fp, dr_fp}) is the name given
to the approximately two-dimensional manifold defined by early-type
galaxies within the three-dimensional space spanned by the logarithms of
effective radius $R_e$, surface brightness $I_e$, and central velocity
dispersion $\sigma_c$.  
When the FP is expressed in the form
\begin{equation}
\label{fpform}
\log_{10} R_e = a \log_{10} \sigma_c + b \log I_e + d~,
\end{equation}
the coefficients $a$ and $b$ are found to be in the
approximate ranges 1.1 to 1.6 and $-0.75$ to $-0.8$ respectively,
depending upon the sample, wave-band, and methods of observation
and fitting (see Tables 2 and~4 of \citealt{bernardi_fp}).
As has been discussed extensively in the
literature, the implications of this
relationship can be understood in terms of a simple
dimensional analysis of the form
\begin{equation}
\label{fpdim}
R_e = c G^{-1} \sigma_c^2 \Upsilon^{-1} I_e^{-1}~,
\end{equation}
where $\Upsilon$ is the total mass-to-light ratio within
some physical aperture and $c$ is a dimensionless constant
parameterizing the details of mass-dynamical structure.
Equation~\ref{fpform} with the values $(a,b) = (2,-1)$ is often
referred to as the ``virial plane'', and the observational fact
that $a < 2$ and $b > -1$ is referred to as the ``tilt'' of the
FP relative to the virial plane.  The terminology is perhaps
misleading: the tilt of the FP does not imply a lack of virial
equilibrium in the individual galaxies that define the plane;
it simply requires that $c$ and/or $\Upsilon$ vary as
a function of the other observables.

\begin{figure*}
\plotone{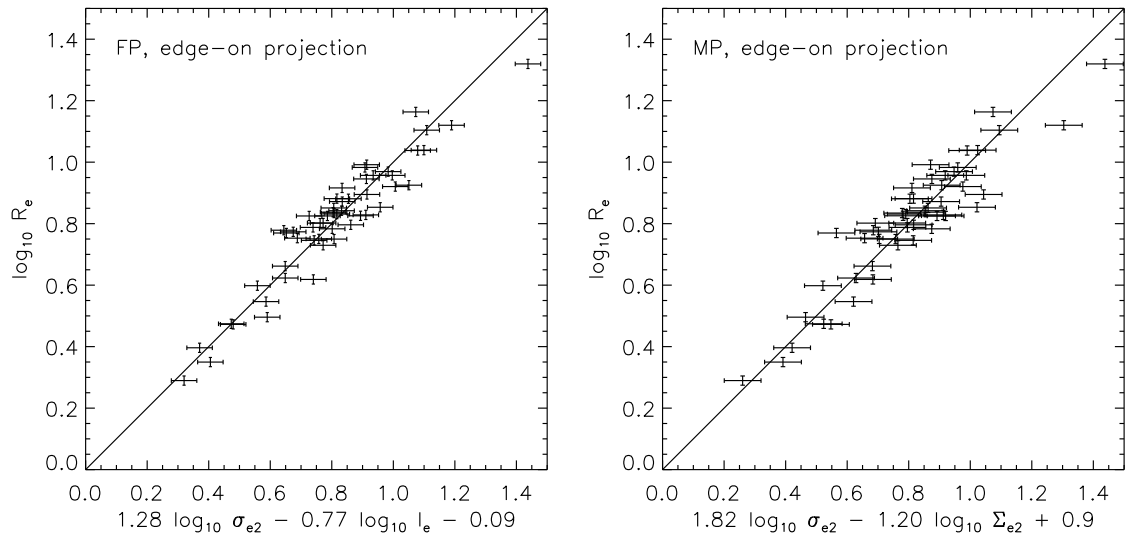}
\caption{\label{fpfig}
Edge-on projections of the best-fit FP (left) and MP
(right) relationships for
the SLACS lens sample. Units are 100\,km\,s$^{-1}$ for $\sigma_{e2}$,
$10^9 L_{V,\odot}$\,kpc$^{-2}$ for $I_e$,
$10^9 M_{\odot}$\,kpc$^{-2}$ for $\Sigma_{e2}$, and
kpc for $R_e$.  These are the same units used in the
FP and MP fits, and were chosen so as to lessen covariance
between plane coefficients and plane zero-point.
Diagonal solid lines represent equality between $\log_{10} R_e$
as observed and as predicted from the best-fit FP\@.
}
\end{figure*}

B07 constructed the FP of SLACS lenses and showed that the
SLACS FP coefficients were consistent with those of the larger
SDSS early-type galaxy population, while Paper~II showed consistency
between the SLACS lenses and the FP of local galaxies, corrected
for luminosity evolution.
We now revisit the calculation of B07
for the larger current sample of lenses (53 early-type
lens galaxies with well-measured SDSS velocity dispersions), using the data
presented in Paper~V\@.\footnote{Our current determination also differs
from that of B07 in the use of luminosities
based upon de-bugged $k$-corrections as described in Paper~V\@.  The
overall conclusions of B07 are unchanged.}
We describe the FP in the form of
Equation~\ref{fpform}, with $\sigma_{e2}$ taking the
role of central velocity dispersion $\sigma_c$.
As noted above, the aperture of $R_e / 2$ is chosen to reflect the
approximate median ratio of Einstein radius $R_{\mathrm{Ein}}$
to $R_e$ within the sample.
Due to the velocity-dispersion aperture correction formula
that we employ \citep{jorgensen_vdisp},
the $\sigma_{e2}$ values are related trivially to the more commonly
quoted $\sigma_{e8}$ values by a fixed factor, with
$\log_{10} \sigma_{e8} = \log_{10} \sigma_{e2} + 0.0241$.
The best-fit FP coefficients
as determined by the
method described in \S\ref{overview}
are presented in Table~\ref{fpcoeffs}.  The plane is
shown in edge-on projection in the left panel of Figure~\ref{fpfig}.
The RMS residual scatter in $\log_{10} R_e$ of the fit is 0.064\,dex.
The RMS error-scaled orthogonal
residual scatter of the fit is 1.44, reasonably consistent with
error estimates but indicating a degree of intrinsic scatter.
The best-fit coefficients are consistent
with those determined by \citet{bernardi_fp} from SDSS data for a
larger early-type galaxy sample, though the details of that analysis
are somewhat different than ours.

As originally noted in B07, there is a significant correlation between
log-radius residuals about the best-fitting FP and the rest-frame $V$-band
mass-to-light ratio $\Upsilon_V$ (for mass and light within
$R_e / 2$, evaluated from B-spline light models and SIE mass models).
This correlation has a linear coefficient $r=-0.337$
(corresponding to a 98.6\% significance) and is in the sense that
lens galaxies with effective radii larger than their FP-predicted
values tend to have lower mass-to-light ratios.  This is to
be expected from a simple dimensional analysis as in Equation~\ref{fpdim}:
if $\sigma$ and $I$ are held fixed (i.e., for a given point on the FP),
an increase in $R$ should give a decrease in $\Upsilon_V$. 
This correlation suggests \textit{observationally}
that the scatter about the FP is due at least in part to
an intrinsic scatter in mass-to-light ratios at
a given point in the plane.

In addition to the FP, we consider the ``mass plane'' (MP) of SLACS lenses
as defined in B07.  By using strong-lensing aperture masses
(corrected to the uniform aperture of $R_e / 2$), we can
replace the surface brightness $I_e$ with the surface mass
density $\Sigma_{e2}$ within $R_e / 2$.
The MP formalism is attractive in that it
is independent of any
luminosity-evolution effects that occur in a spatially uniform manner within
the lens galaxies.  Thus the MP can be tracked across redshift without
correction for the dimming of stellar luminosity.
The MP should also be better suited
to comparison with theory and numerical simulation of galaxy formation,
merging, and dynamical evolution processes.
Since the scale length of the MP is still taken
from the luminosity distribution and the MP velocities are still traced
by the stars, it is not a pure mass space formulation.
However, one can extract total (luminous plus dark matter)
aperture masses, stellar effective radii, and stellar
velocity dispersions from any sensibly constructed theory of galaxy
formation, evolution, and merging.  One cannot extract luminosities
without delving into the entirely different domain of stellar
populations and their evolution.
Thus the MP formulation---and more elementally,
the data upon which it is based---represent
an important step in bringing theory and observation together by separating
the subject of stellar populations from the subject of galaxy formation.

We express the MP in a form analogous to the FP:
\begin{equation}
\label{mpform}
\log_{10} R_e = a_m \log_{10} \sigma_{e2} + b_m \log_{10} \Sigma_{e2} + d_m~.
\end{equation}
Since $\Sigma_{e2}$ depends somewhat upon the lens model
used to evaluate the surface density, we fit for MP coefficients using
$\Sigma_{e2}$ values from both the SIE (singular isothermal
ellipsoid) and LTM (light-traces-mass) mass models
of Paper~V\@.  Values for the best-fit coefficients are given in Table~\ref{fpcoeffs}.
The best-fit MP (using the SIE aperture masses)
is seen in edge-on projection in the right-hand panel of Figure~\ref{fpfig}. %,
The RMS residual scatter in $\log_{10} R_e$ about the best-fit MP is
0.076\,dex using SIE surface densities and 0.10\,dex using LTM surface densities.
The RMS error-scaled orthogonal residuals are 1.24 (SIE) and 1.50 (LTM).
For the case of the SIE-based MP, the plane is tighter than the
FP when judged by the RMS error-scaled orthogonal residuals
(the quantity minimized in the fitting), but not exceptionally so:
the edge-on MP is not noticeably tighter than the edge-on FP
as seen in of Figure~\ref{fpfig}.

We can see from Table~\ref{fpcoeffs} that the
coefficients of the MP are in a sense less ``tilted'' relative
to the values $(a_m,b_m) = (2,-1)$.  To quantify this, we consider
$10^6$ bootstrap re-samplings of our lens sample, from which we compute
confidence limits in the $a_m$--$b_m$ plane.  The values $(a_m,b_m) = (2,-1)$
fall on the contour enclosing 77\% (60\%) of the
re-sample coefficient points for SIE (LTM) based fits,
showing that the MP is in fact fairly consistent with the
plane represented by $(a_m,b_m) = (2,-1)$.
We cast this statement in more physical terms in \S\ref{mml}.

\begin{table}
\begin{center}
\caption{\label{fpcoeffs} FP and MP coefficients}
\begin{tabular}{cccc}
\hline \hline
Surface &  ~  &  ~  &  ~  \\
Density & $a$ & $b$ & $d$ \\
\hline
$I_{e}$ & $1.28 \pm 0.22$ & $-0.77 \pm 0.07$ & $-0.09 \pm 0.07$ \\
$\Sigma_{e2,\mathrm{SIE}}$ & $1.82 \pm 0.19$ & $-1.20 \pm 0.12$ & $\phn 0.91 \pm 0.14$ \\
$\Sigma_{e2,\mathrm{LTM}}$ & $2.10 \pm 0.24$ & $-0.83 \pm 0.12$ & $\phn 0.56 \pm 0.15$ \\
\hline
\end{tabular}
\tablecomments{Coefficients are for FP and MP relations in the form
$\log_{10} R_e = a \log_{10} \sigma_{e2} + b \log_{10} \mbox{(surface density)} + d$.
$R_e$ measured in kpc, $\sigma_{e2}$ in units of 100\,km\,s$^{-1}$,
$I_e$ in units of $10^9 L_{V,\odot}/$\,kpc$^{-2}$,
and $\Sigma_{e2}$ in units of $10^9 M_{\odot}/$\,kpc$^{-2}$.
Units are chosen so as to reduce artificial covariance between plane
coefficients and plane zero-points.
Fit method is described in \S\ref{overview}.}
\end{center}
\end{table}

\section{Lensing mass, dynamical mass, and luminosity}
\label{mml}

Depending on the details of the coefficients,
the FP can be understood in terms of
a systematic relation between luminosity and a dimensional
mass variable\footnote{Our ``dimensional mass'' $M_{\mathrm{dim}}$ is, modulo
a constant factor, equal to the ``effective mass'' or ``dynamical mass'' referred
to by other authors.  We use the term ``dimensional'' to avoid confusion.}
such as $M_{\mathrm{dim}} \equiv G^{-1} \sigma_{e2}^2 (R_e / 2)$
of the form $L \propto  M_{\mathrm{dim}}^{\eta}$
\citep[e.g.,][]{faber_87, bbf_92, clr_fp_96}.  Such a relation
leads to FP coefficients in Equation~\ref{fpform} given
by $a = 2 \eta / (2 - \eta)$ and $b = -1 / (2 - \eta)$.
This may be conceptualized as a systematic variation of a
``dimensional mass-to-light ratio''
$\Upsilon_{\mathrm{dim}} \equiv M_{\mathrm{dim}} / L$ with
$M_{\mathrm{dim}}$.  It is, however, important to keep in mind that
$\Upsilon_{\mathrm{dim}}$ is linearly
proportional to the \textit{true} mass-to-light ratio $\Upsilon$
only if the FP tilt is due to a systematic mass-to-light ratio trend,
rather than to a trend in mass-dynamical structure.

The availability of strong-lensing aperture masses
in addition to the traditional FP observables allows us to
directly test the alternative hypotheses for the ``tilt'' of
the FP, because in addition to $M_{\mathrm{dim}}$, we measure the
aperture mass $M_{\mathrm{lens}}$ from strong lensing.
Consider the following two relations:
\begin{eqnarray}
\label{mlformone}
L_V &=& c_1 M_{\mathrm{dim}}^{\eta}~, \\
\label{mlformtwo}
M_{\mathrm{lens}} &=& c_0 M_{\mathrm{dim}}^{\delta}~.
\end{eqnarray}
Consider also the following two alternative hypotheses:
(1) on average, early-type galaxies have a universal mass-dynamical
structure, and the tilt of the FP is due to a systematic trend in
total mass-to-light ratio with mass; or (2) on average,
early-type galaxies have a universal mass-to-light ratio, and the
tilt of the FP is due to a systematic trend in mass-dynamical
structure.  If hypothesis (1) is correct, then
we should find $\delta \simeq 1$, whereas if hypothesis (2)
is correct, we should find $\delta \simeq \eta$.
Of course, we may also find that $\eta < \delta < 1$,
since in principle both explanations could contribute to the
tilt of the FP\@.  Roughly speaking, hypothesis (1) represents ``homology''
and hypothesis (2) represents ``non-homology''.

Before proceeding, we note that
numerous works have explored the possible role of a systematically
varying S\'{e}rsic index $n$ in causing the tilt of the FP
\citep[e.g.,][]{hjorth_madsen, graham_colless_97, bcdp_02, tbb_fp}, and
thus we must not disregard this possibility in our own analysis.
We compute $n$ for all SLACS lenses
by continuing the de Vaucouleurs model optimizations
after freeing the index from its fixed $n = 1/4$ value.
We find that $n$ is completely uncorrelated with
lensing mass, dynamical mass, luminosity, and velocity dispersion
within the sample, and therefore the inclusion of the S\'{e}rsic
index as a significant factor in our analysis is not
motivated by the data.  This lack of correlation
between $n$ and other quantities is in fact consistent
with other studies, since the SLACS sample is confined
to relatively high-mass/high-luminosity early-type galaxies,
and does not extend over a sufficient range to define
the $n$--$L$ correlation significantly given the level of intrinsic
scatter \citep[e.g.,][]{caon_93, donofrio_94, graham_guzman, ferrarese_06}.

\begin{figure*}
\plotone{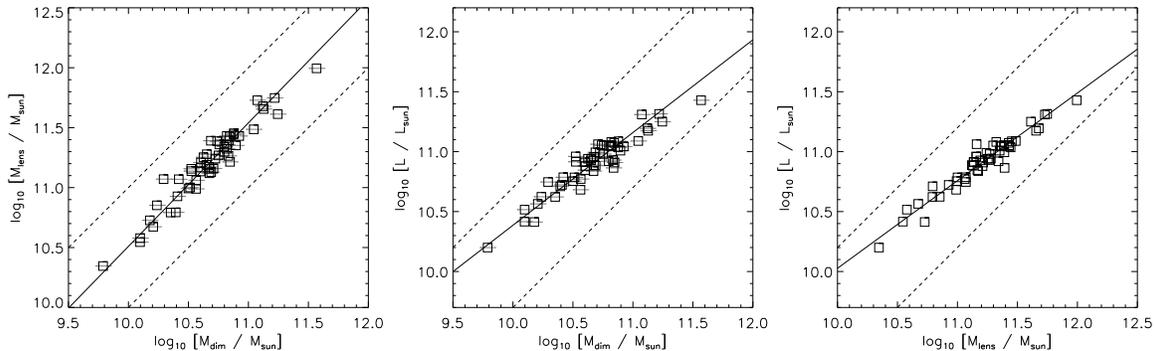}
\caption{\label{m_vs_l}
Relationships between mass and luminosity variables.  \textsl{Left panel:}
Lensing mass $M_{\mathrm{lens}}$ within one-half effective radius (evaluated
from the SIE lens models) versus
dimensional mass variable $M_{\mathrm{dim}} = G^{-1} \sigma_{e2}^2 (R_e / 2)$.
\textsl{Center panel:} Rest-frame $V$-band luminosity $L$ versus $M_{\mathrm{dim}}$.
\textsl{Right panel:} Rest-frame $V$-band luminosity $L$ versus $M_{\mathrm{lens}}$.
In each panel, the best-fit linear relationship between the two logarithmic quantities
is shown by the solid line.
Dashed lines indicate a slope of unity, for reference.
Typical errors in $M_{\mathrm{dim}}$ are shown by
gray error bars in the left and center panels; errors in $M_{\mathrm{lens}}$
and $L$ are smaller than the plot symbols.}
\end{figure*}

Table~\ref{mlcoeffs} shows the results of
fits for the normalizations and exponents of the
relations defined in Equations~\ref{mlformone} and~\ref{mlformtwo},
as well as for the relation between luminosity and lensing
mass.  Within the uncertainties, the clear result is that
$\delta \simeq 1$ while $\eta < 1$: thus, we conclude based upon
our sample of lenses that the tilt of the FP---as defined by
massive ellipticals---is due to a
systematic trend in mass-to-light ratio and not to a trend
in mass-dynamical structure.
These relations are shown graphically in Figure~\ref{m_vs_l},
where we can see by eye that the logarithmic slope of the
$M_{\mathrm{lens}}$-versus-$M_{\mathrm{dim}}$ relation is
significantly less shallow than those of
the $L_V$-versus-$M_{\mathrm{dim}}$ and $L_V$-versus-$M_{\mathrm{lens}}$ relations.
The slopes of the $L$-versus-$M$ relations change
only negligibly when the de Vaucouleurs model luminosity is replaced with
the more general radial B-spline model aperture luminosity within $R_e / 2$
(described in Paper~V).
If we fix $\delta = 1$, we find
$\left< \log_{10} [M_{\mathrm{lens}} / M_{\mathrm{dim}}] \right> = 0.530 \pm 0.012$
(RMS of 0.08) for SIE aperture corrections and
$\left< \log_{10} [M_{\mathrm{lens}} / M_{\mathrm{dim}}] \right> = 0.543 \pm 0.014$
(RMS of 0.1) for LTM corrections.  Accounting for our best estimates of the
typical measurement errors, the intrinsic scatter in
$\log_{10} [M_{\mathrm{lens}} / M_{\mathrm{dim}}]$ about the mean value
that gives a reduced $\chi^2$ of approximately unity
is 0.057\,dex RMS, or about a $\pm$13\%.  This is consistent with the
intrinsic velocity-dispersion scatter of 7.5\% seen in \S\ref{v_v},
given that $M_{\mathrm{dim}}$ scales as the square of velocity dispersion.

\begin{table*}
\begin{center}
\caption{\label{mlcoeffs}Mass and light scaling relations}
\begin{tabular}{lcc}
\hline \hline
Scaling Relation  &  Prefactor  &   Exponent \\
\hline
  $(M_{\mathrm{SIE}}/10^{11} M_{\odot}) = c_0 (M_{\mathrm{dim}}/10^{11} M_{\odot})^\delta$
   & $\log_{10} c_0 =  0.54 \pm 0.02$ & $\delta = 1.03 \pm 0.04$ \\
  $(M_{\mathrm{LTM}}/10^{11} M_{\odot}) = c_0 (M_{\mathrm{dim}}/10^{11} M_{\odot})^\delta$
   & $\log_{10} c_0 =  0.54 \pm 0.02$ & $\delta = 0.99 \pm 0.05$ \\
  $(L_V/10^{11} L_{\odot}) = c_1 (M_{\mathrm{dim}}/10^{11} M_{\odot})^\eta$
   & $\log_{10} c_1 =  0.16 \pm 0.02$ & $\eta = 0.77 \pm 0.04$ \\
  $(L_V/10^{11} L_{\odot}) = c_2 (M_{\mathrm{SIE}}/10^{11} M_{\odot})^{\eta\prime}$
   & $\log_{10} c_2 = -0.24 \pm 0.01$ & $\eta^{\prime} = 0.73 \pm 0.03$ \\
  $(L_V/10^{11} L_{\odot}) = c_2 (M_{\mathrm{LTM}}/10^{11} M_{\odot})^{\eta\prime}$
   & $\log_{10} c_2 = -0.26 \pm 0.01$ & $\eta^{\prime} = 0.77 \pm 0.03$ \\
\hline
\end{tabular}
\tablecomments{Mass values $M_{\mathrm{SIE}}$ and $M_{\mathrm{LTM}}$
are masses within $R_e/2$ as evaluated from SIE and LTM lens models.
Dimensional mass variable is defined as
$M_{\mathrm{dim}} \equiv G^{-1} \sigma_{e2}^2 (R_e / 2)$.
Fit method is described in \S\ref{overview}.}
\end{center}
\end{table*}

\begin{figure}
\plotone{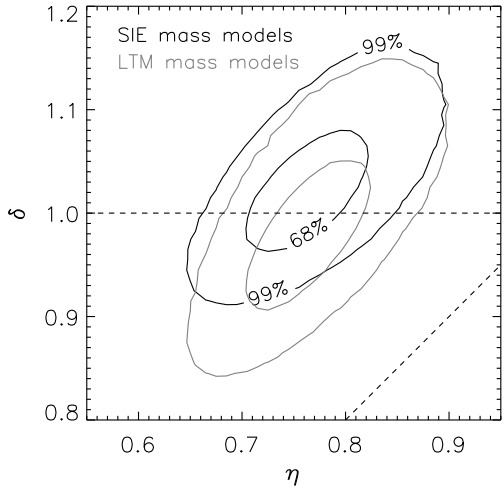}
\caption{\label{signif_plot}
Distribution of $\eta$ and $\delta$ parameter fits
to $10^6$ bootstrap re-samples.  Results are shown for lensing
aperture masses evaluated using both SIE lens models (black) and LTM lens models (gray).
Dashed lines of $\delta = 1$ and of
$\delta = \eta$ are shown as well.}
\end{figure}

The result that $\delta \simeq 1$ while $\eta < 1$ is one of the central results
of this work and of B07,
and thus its significance merits
special attention.
Specifically, we consider the distribution of
pairs of $\eta$ and $\delta$ values \textit{fitted to the same sets of
bootstrap samples}.  For both the SIE and LTM aperture masses, we generate $10^6$
bootstrap re-samples and fit $\eta$ and $\delta$ for each.
The resulting distributions can be seen in Figure~\ref{signif_plot}.
We see that for both mass models,
$\delta$ is significantly greater that $\eta$ and consistent with $\delta = 1$,
while $\eta = 1$ and $\delta = \eta$ are ruled out at high significance.

The skeptic might worry that our result of a universal mass-density structure
as a function of mass has been ``baked in the cake'' by our
use of a single universal profile (i.e., the isothermal model or the approximately
de Vaucouleurs LTM model) to evaluate aperture masses for
the uniform physical aperture of $R_e / 2$.  The robustly measured
quantity is the mass within the Einstein radius $R_{\mathrm{Ein}}$
\citep{kochanek_91}, and in general $R_{\mathrm{Ein}}$ differs
from $R_e / 2$ (though, by design, not by a large amount).
If such an explanation were masking a true mass-dependent trend
in structure, we would expect to see a correlation of $R_{\mathrm{Ein}} / R_e$
with the residuals about the best-fit $M_{\mathrm{lens}}$-versus-$M_{\mathrm{dim}}$ 
relationship.  In fact, we see no such correlation.  In addition,
if we restrict our analysis to that half of the lens sample for which
$R_{\mathrm{Ein}}$ is most nearly matched to $R_e / 2$
(as quantified by the fractional difference between the aperture masses
within $R_{\mathrm{Ein}}$ and $R_e / 2$),
we see no significant changes in our results.

We can easily translate our relations between $V$-band luminosity and
lensing mass within $R_e / 2$ (the last two rows of Table~\ref{mlcoeffs})
into expressions for the mass-to-light
ratio within $R_e/2$ as a function of luminosity.
Using the fact that
32.0\% of the total de Vaucouleurs model flux is enclosed within
the $R_e / 2$ aperture, we have
\begin{equation}
{{\Upsilon_{V,e2}} \over {\Upsilon_{V,\odot}}} = 
c_2^{-1/\eta\prime} \left( {1 \over {0.320}} \right)
\left( {{L_V} \over {L_{V,\odot}}} \right)^{ [(1 / \eta\prime) - 1] }~.
\end{equation}
For SIE aperture masses, we find
$\log_{10} (\Upsilon_{V,e2}/\Upsilon_{V,\odot}) =
(0.83 \pm 0.01) + (0.37 \pm 0.06) \log_{10} (L_V/10^{11}L_{V,\odot})$,
while for LTM aperture masses we obtain
$\log_{10} (\Upsilon_{V,e2}/\Upsilon_{V,\odot}) =
(0.83 \pm 0.01) + (0.29 \pm 0.05) \log_{10} (L_V/10^{11}L_{V,\odot})$---i.e.,
a total central $V$-band mass-to-light ratio of approximately
6.8 times solar for a $V$-band luminosity of $10^{11}$ times solar.
This $\Upsilon$--$L$ relation should not necessarily
be regarded as fundamental, however: we find
comparably significant correlations of $\Upsilon$
with $M_{\mathrm{lens}}$, $M_{\mathrm{dim}}$,
$R_e$, and $\sigma_{e8}$.

The details of our mass-luminosity relations
are sensitive to the assumed rate of passive evolution which we have attempted
to remove from the sample.  The SLACS lens sample exhibits a significant
luminosity--redshift degeneracy (see Paper~V) which prevents us from solving
simultaneously for the mass and redshift dependence of luminosity.
The main effect that we expect is a certain level of unmodeled luminosity
dependence in the rate of $V$-band evolution, characteristic of
more recent star formation (and hence faster fading) in less luminous galaxies
\citep[e.g.,][]{cowie_96, treu05a, treu05b, vanderwel_06},
though the effect should be less
pronounced in the $V$ band than in the $B$ band.
In the absence of published constraints on the
mass dependence of $V$-band luminosity evolution,
we consider the effects of a simple ``toy model''
whereby the rate of luminosity evolution
varies from $d \log_{10} L_V / dz = 0.6$ to 0.2
linearly over the range
$\log_{10} (M_{\mathrm{lens}} / M_{\odot}) = 10$ to $12$.
(This is likely to be an extreme scenario---see the rest-frame $B$-band
analysis of \citealt{treu05b}.)
In this case, the exponents of the
$L$-versus-$M$ relations given in Table~\ref{mlcoeffs}
change from $\eta = 0.77$ to $\eta = 0.82$ and
$\eta^{\prime} = 0.73$ to $\eta^{\prime} = 0.78$,
with unchanged uncertainties.  We defer more detailed analysis until the
completion of multi-band observations
currently underway with \textsl{HST}-WFPC2, which will enable the determination
of rest-frame $B$-band luminosities and subsequent connection to a more
comprehensive literature on luminosity evolution
\citep[e.g,][]{treu_2002, treu05b, vanderwel_05, dsa_05}.
These ambiguities of
luminosity evolution have no bearing upon the pure mass-dynamical relations,
and no effect upon our result that the structure of massive elliptical galaxies
is (on average) independent of galaxy mass.

\section{Robust ensemble measurement of the radial mass-density profile}
\label{mofr}

\begin{figure}
\plotone{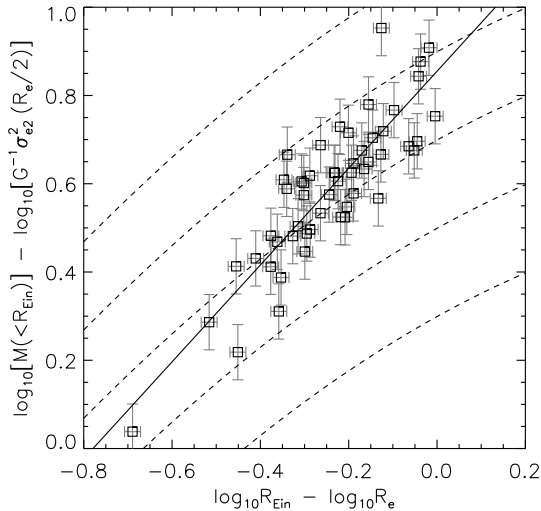}
\caption{\label{massrad}
Dimensionless projected mass-radius relation.  Horizontal axis
is mass-aperture (Einstein) radius scaled by lens-galaxy
effective radius.  Vertical axis
is lensing aperture mass within the Einstein radius scaled by the
dimensional mass variable formed from the combination of
central velocity dispersion and effective radius.  Error bars show
the typical error in each coordinate used in the scaled
orthogonal fit for the best relation.  The solid line shows this
best-fit linear relation in the logarithmic coordinates, given by
$\log_{10} [M(<R) / M_{\mathrm{dim}}] = 
(1.10 \pm 0.09) \log_{10} [R / R_e]
+ (0.85 \pm 0.03)$.  For comparison, dashed lines show the logarithmic
enclosed-luminosity profile for the de Vaucouleurs
surface-brightness model, with various offsets.  If a light-traces-mass
model were correct, the data points would trace the slope of
these de Vaucouleurs curves.}
\end{figure}

In this section, we constrain
the average projected mass-density profile of
our lens galaxies without any dynamical modeling by considering
the lens ensemble as a whole,
and using the fact that the aperture masses are
measured most robustly for physical apertures (i.e., Einstein
radii) ranging from $0.2$ to 1 times
the effective radius depending upon the
individual lens.\footnote{In fact,
the sample probes three-dimensional radii even beyond the physical
scale of $R_e$, due to the sensitivity of lensing
to projected mass.  For the singular isothermal sphere,
36\% of the mass (the exact fraction is $1 - 2/\pi$)
within a cylinder of radius $R$ is exterior
to the corresponding sphere of the same radius.}  Motivated by the
FP, we regard the sample as a two-parameter family,
parameterized by effective radius $R_e$ and velocity dispersion $\sigma_{e2}$.
From these quantities, we construct the previously defined
``dimensional mass'' $M_{\mathrm{dim}} \equiv G^{-1} \sigma_{e2}^2 (R_e / 2)$.
We then scale the measured aperture masses
within the Einstein radii $M(<\!\!R) = M_{\mathrm{Ein}}$ by
$M_{\mathrm{dim}}$, and scale the corresponding apertures $R = R_{\mathrm{Ein}}$
by the effective radii $R_e$.  The resulting dimensionless
projected mass-radius relation is shown in
logarithmic coordinates in Figure~\ref{massrad}.
Describing this relation with a power law (i.e., linear
in the logarithmic coordinates) in the form
\begin{equation}
\log_{10} [M(<\!\!R) / M_{\mathrm{dim}}] = 
g \log_{10} [R / R_e] + h~,
\end{equation}
we find $g = 1.10 \pm 0.09$ and $h = 0.85 \pm 0.03$ for SIE
aperture masses or $g = 1.13 \pm 0.11$ and $h = 0.85 \pm 0.03$ for LTM
aperture masses.
RMS orthogonal residuals scaled by error estimates are
1.27 (SIE) and 1.41 (LTM), indicating reasonable consistency of
the data with the fitted relation, but with
evidence for intrinsic scatter as in other relations.
The residuals about the best-fit relation are completely
uncorrelated with $M_{\mathrm{dim}}$, an echo
of the mass-independent structure result of \S\ref{mml}.
For a pure singular isothermal model,
the projected aperture mass scales as
$M(<\!\!R) \propto R$---i.e., $g=1$.
Thus we find that our data are consistent with
the universal total-mass profile being isothermal.
Translating back into a three-dimensional power-law profile,
our results correspond to $\rho(r) \propto r^{-1.90 \pm 0.09}$
(using SIE aperture masses)
and $\rho(r) \propto r^{-1.87 \pm 0.10}$
(using LTM aperture masses).
This analysis is very similar to that employed
by \citet{rkk_03} and \citet{rusin_kochanek_05} to derive
constraints on the quasar-lens population.
The key difference for our application to SLACS is that,
since we have stellar velocity dispersions for our
entire sample, we can directly scale lensing masses by dynamical
masses, ignoring issues of luminosity trends and evolution rates.
The most likely average power-law profile found by \citet{rusin_kochanek_05}
is slightly steeper ($\rho(r) \propto r^{-2.06 \pm 0.17}$) than that
found in this work, but the two results are consistent within their
combined errors.  Our (nearly) isothermal result is also consistent with the
lens$+$dynamical modeling results of LSD and Paper~III,
with the dynamical analysis of \citealt{gerhard_2001}, and with
the strong$+$weak lensing analysis of Paper~IV\@.
The particular advantage of the method
presented here is that it is simple and
robust, with little susceptibility to assumptions or systematic errors.
(e.g., \citealt{gerhard_2001}; \citealt{rkk_03}; LSD; Paper~III; Paper~IV\@).

For comparison, the logarithmic slope of the projected
enclosed \textit{luminosity}
profile changes very little over the range 0.3 to 1 times $R_e$ where
the bulk of our data are concentrated, having a value
of $g_{\mathrm{light}} = 0.71$ for the de Vaucouleurs model and a mean value
of $g_{\mathrm{light}} = 0.68$ for the more
general B-spline luminosity profile models of Paper~V\@.
If a light-traces-mass model were appropriate, we would
expect to find similar values for the logarithmic slope of
enclosed mass with radius.
The one-sided 99\%, 99.9\%, and 99.99\% lower limits on the
mass slope variable $g$ from bootstrap re-samples are
0.85, 0.76, and 0.67 respectively for SIE fits
(0.83, 0.72, and 0.62 for LTM fits).
Thus with only three basic observables---velocity dispersion,
effective radius, and Einstein radius (as translated
into physical units and enclosed mass using
spectroscopic redshifts)---and no dynamical
modeling, we can falsify the
light-traces-mass hypothesis at 99.9\% to 99.99\%
confidence \textit{even inside the effective
radius}, where dark matter is often assumed to play little
role.  This result argues for a preference of SIE
lens model parameters over those from the LTM models.
Figure~\ref{massrad} shows the slope of the de Vaucouleurs
aperture-luminosity profile for comparison with the
enclosed-mass data points; the inconsistency
of the data with an LTM model is visually apparent.

We can translate our measurement into a lower limit on the
average central dark-matter fraction by making a ``maximal bulge''
assumption---specifically, assuming that the total mass and
the stellar mass are equal within $0.3 R_e$.  We then take the
projected stellar mass-profile to go as
$M_{\mathrm{stars}}(<\!\!R) \propto R^{0.7}$ (following the
luminosity profile) and the projected total-mass profile
to go as $M_{\mathrm{total}}(<\!\!R) \propto R^{1.1 \pm 0.1}$.
This gives a lower limit on the
average \textit{projected} dark-matter fraction
of $f_{\mathrm{DM}} = 0.38 \pm 0.07$ within one
effective radius $R_e$.  This is consistent with the
highest values found through dynamical analysis of
nearby elliptical galaxies \citep[e.g.,][]{gerhard_2001}.

\section{Summary of results}
\label{summary}

By combining the direct mass measurements provided by the
strong gravitational lensing effect with traditional
measures of galaxy luminosity, size, and velocity dispersion,
we have derived empirical scaling laws for massive early-type
galaxies.  The results of our analyses are summarized as follows.
We remind the reader that these results apply specifically
to galaxies with
velocity dispersions between approximately 175 and 400\,km\,s$^{-1}$,
rest-frame $V$-band luminosities between $10^{10}$ and
$3 \times 10^{11} L_{\odot}$, and redshifts between
0.06 and 0.36.  (The few higher redshift SLACS lenses presented
in Paper~V have an SDSS spectral SNR that is too low for a reliable
velocity-dispersion measurement.)

\begin{enumerate}

\item The projected major axes of the mass and light
distributions within the SLACS lens sample are aligned to
$\left< \Delta PA \right> = 2.5^{\circ} \pm 2.4^{\circ}$ (RMS of $18^{\circ}$) for
55 lenses with light axis ratios $q_{\mathrm{stars}} < 0.95$.  Restricted to
the subset of 29 lenses whose angular structure is tightly
constrained by lensed images through an extended azimuthal range
about the lens center, this alignment tightens to
$\left< \Delta PA \right> = 1.4^{\circ} \pm 1.9^{\circ}$, with an RMS
of $10^{\circ}$.

\item For light-traces-mass (LTM) lens models, the required external
shears are preferentially aligned with the major axes of the lens
galaxies at extremely high significance when the sample as a whole
is considered.  This alignment suggests a falsification
of the LTM hypothesis.

\item The average relative flattening between mass and light
for the SLACS sample, measured by the ratio of projected
(2D) minor-to-major axis ratios $q_{\mathrm{SIE}}$ (mass)
and $q_{\mathrm{stars}}$ (light), is $\left< q_{\mathrm{SIE}} / q_{\mathrm{stars}} \right>
= 1.02 \pm 0.02$ (RMS of 0.12): i.e., consistent with
unity on average.  We see a decreasing
trend in $q_{\mathrm{SIE}} / q_{\mathrm{stars}}$ with increasing velocity
dispersion $\sigma_{\mathrm{SIE}}$ (measured from the
SIE lens models).

\item The ratio $f \equiv \sigma_{e8} / \sigma_{\mathrm{SIE}}$
of central stellar velocity
dispersions to isothermal lens-model velocity dispersions
is $f = 1.019 \pm 0.008$.
The RMS scatter about the mean is approximately 0.1, which
when considered along with measurement errors corresponds to
an intrinsic velocity scatter of about 20\,km\,s$^{-1}$
(or about 7.5\% of $\sigma_{\mathrm{SIE}}$).
The residual scatter about the mean relation is not
correlated with mass, luminosity, velocity dispersion,
mass-to-light ratio, effective radius, or ratio
of Einstein radius to effective radius.

\item The SLACS lens galaxies define a fundamental plane (FP)
that is consistent with the FP measured for the general population
of early-type galaxies observed by the SDSS and other surveys.

\item The SLACS lens sample also defines
a ``mass plane'' (MP) relation, obtained by replacing
surface brightness with surface mass density as measured from
strong lensing.  The MP is significantly
less ``tilted'' than the FP, and is essentially consistent with
simple expectations based on the virial theorem and a universal
mass-dynamical structure within the population.

\item \label{ml_item} The relationship between rest-frame $V$-band
luminosity $L_V$ and the dimensional mass variable
$M_{\mathrm{dim}} \equiv G^{-1} \sigma_{e2}^2 (R_e / 2)$
is given by
$\log_{10} [L_V/10^{11} L_{\odot}] = (0.77 \pm 0.04) \log_{10}
[M_{\mathrm{dim}}/10^{11} M_{\odot}] + (0.16 \pm 0.02)$.
A similar relation holds when $M_{\mathrm{dim}}$ is replaced
with the strong lensing-determined mass within $R_e/2$:
$\log_{10} [L_V/10^{11} L_{\odot}] = (0.73 \pm 0.03) \log_{10}
[M_{\mathrm{lens},e2}/10^{11} M_{\odot}] - (0.24 \pm 0.01)$.
The consistency between the slopes of these two relations implies that the
systematic ``dynamical'' mass-to-light variation inferred from
the FP is representative of a \textit{true} mass-to-light
variation within the early-type galaxy population.

\item \label{mm_item} The relationship between $M_{\mathrm{dim}}$
and the mass $M_{\mathrm{lens}}$ within $R_e / 2$ as measured by strong lensing
is given by $\log_{10} [M_{\mathrm{lens}}/10^{11} M_{\odot}]
= (1.03 \pm 0.04) \log_{10} [M_{\mathrm{dim}}/10^{11} M_{\odot}]
+ (0.54 \pm 0.02)$.
The unitary (within errors) slope of
this $M$-versus-$M$ relation indicates that
the mass-dynamical structure
of early-type galaxies does not vary systematically with mass
over the range probed by the SLACS sample
(approximately $9.8 < \log_{10} [M_{\mathrm{dim}} / M_{\odot}] < 11.6$). %,
Our result indicates that the dimensional mass is a
suitable proxy for the true mass within the central regions
of massive early-type galaxies.
Fitting directly for an overall scaling between
lensing and dimensional masses, we find that
$\left< \log_{10} [M_{\mathrm{lens}} / M_{\mathrm{dim}}] \right> = 0.530 \pm 0.012$
with an RMS scatter of 0.08\,dex about the mean.
After accounting for measurement errors, this scatter is
approximately 0.057\,dex (or about $\pm$13\%).

\item The difference between the slopes of the $M$--$L$ relations of
result~\ref{ml_item} and the $M$--$M$ relation of result~\ref{mm_item}
is very highly significant, and implies that the ``tilt'' of the
FP is due to a systematic variation in the total mass-to-light
ratio with mass or luminosity, rather than to a systematic
variation in mass-dynamical structure.  The SLACS sample shows no
significant correlation between luminosity and S\'{e}rsic index $n$.

\item Expressed as a function of luminosity,
the central $V$-band total mass-to-light ratio $\Upsilon_{V,e2}$ of SLACS lenses
scales according to $\log_{10} (\Upsilon_{V,e2}/\Upsilon_{V,\odot} =
(0.83 \pm 0.01) + (0.37 \pm 0.06) \log_{10} (L_V/10^{11}L_{V,\odot})$.

\item We obtain a nearly model-independent
ensemble constraint on the average mass-density profile
in the SLACS sample by assuming (in accordance with the FP)
that the lenses form a two-parameter family.
We scale strong-lensing aperture masses by $M_{\mathrm{dim}}$
and radial mass apertures (Einstein radii) by $R_e$, and find
a non-dimensional projected mass-radius relation given by
$\log_{10} [M(<\!\!R) / M_{\mathrm{dim}}] =
(1.10 \pm 0.09) \log_{10} [R/R_e] + (0.85 \pm 0.03)$.
This result is inconsistent with the steeper slope of the projected
aperture luminosity profile at a
level of significance between 99.9\% and 99.99\%,
thus falsifying the light-traces-mass hypothesis.
If we assume that all projected mass interior to
$0.3 R_e$ is in the form of stars---a ``maximal bulge''
assumption---this result translates
into a lower limit on the average
\textit{projected} dark-matter fraction of
$f_{\mathrm{DM}} = 0.38 \pm 0.07$ inside one effective radius.
The three-dimensional mass-density profile corresponding to our
two-dimensional result
is $\rho(r) \propto r^{-1.90 \pm 0.09}$, consistent with the
isothermal (flat rotation curve) model.

\end{enumerate}

\section{Discussion and Conclusions}
\label{discuss}

The SLACS lens sample provides
a unique resource for the quantitative study
of the mass-dynamical structure of massive early-type galaxies.
This is due to the addition of strong-lensing aperture masses to
a full complement of traditional galaxy observables
over a significant range of intrinsic parameter variation.
The fact that the SLACS gravitational lens sample
defines a fundamental plane similar to that defined
by the larger sample of SDSS early-type galaxies
suggests that deductions based upon the SLACS
lenses can be generalized to early-type galaxies
in general.  This conclusion is further
supported by the FP analysis of Paper~II, and by the
analysis in Paper~V of the distribution of SLACS lenses in
luminosity within their parent samples.

\subsection{The mass--velocity connection}

The lack of correlation between
$f \equiv \sigma_{\mathrm{e8}} / \sigma_{\mathrm{SIE}}$
and either mass or $R_{\mathrm{Ein}} / R_e$ in fact contains the essence
of our results on the mass independence of galaxy structure and the
near isothermal nature of the radial profile, as we now illustrate.
Consider the following
relation for the mass enclosed within the Einstein radius,
which is a consequence of lensing geometry and holds for all
mass density models \citep[e.g.,][]{nb96}:
\begin{equation}
\label{mein}
M(R_{\mathrm{Ein}}) = M(bD_{\mathrm{L}}) = {{c^2 b^2} \over {4 G}}
{{D_{\mathrm{L}}D_{\mathrm{S}}} \over {D_{\mathrm{LS}}}}~.
\end{equation}
Furthermore, take Equation~\ref{b_sie} as the definition of
the lens-model velocity dispersion parameter $\sigma_{\mathrm{SIE}}$
in terms of the observable angular Einstein radius $b$
(which is in turn related to the physical Einstein radius
through $R_{\mathrm{Ein}} = b D_{\mathrm{L}}$).
Now consider an idealized case where
$R_{\mathrm{Ein}} = R_e / 2$
across a range of masses.  Equation~\ref{mein} becomes
\begin{eqnarray}
\nonumber
M(R_e/2) &=& \pi \left( {{c^2 b} \over {4 \pi}}
{{D_{\mathrm{S}}} \over {D_{\mathrm{LS}}}} \right)
\left( {{R_e / 2} \over {G}} \right) \\
\nonumber
&=& \pi (\sigma_{\mathrm{SIE}}^2) \left( {{M_{\mathrm{dim}}}
\over {\sigma_{\mathrm{stars}}^2}} \right) \\
&=& \pi f^{-2} M_{\mathrm{dim}}~.
\end{eqnarray}
Thus the lack of a correlation of $f$ with $M_{\mathrm{dim}}$
implies a linear relationship between $M(R_e / 2)$ and $M_{\mathrm{dim}}$.
Similarly, if we now consider
$R = R_{\mathrm{Ein}}$ for any fraction of $R_e$
and divide Equation~\ref{mein} by $M_{\mathrm{dim}}$, we obtain
\begin{eqnarray}
\nonumber
{{M (R)} \over {M_{\mathrm{dim}}}} &=& 
\pi \left( {{c^2 b} \over {4 \pi}}
{{D_{\mathrm{S}}} \over {D_{\mathrm{LS}}}} \right)
\left( {{R} \over {G}} \right) \left( {{G} \over {\sigma_{\mathrm{stars}}^2 R_e / 2}} \right) \\
\nonumber
&=& 2 \pi \sigma_{\mathrm{SIE}}^2 [R / (\sigma_{\mathrm{stars}}^2 R_e)] \\
&=& 2 \pi f^{-2} (R / R_e)~.
\end{eqnarray}
Thus a non-isothermal mass profile, in which the enclosed mass does not
grow linearly with radius, would appear as a correlation of $f$ with
the aperture-radius ratio $R / R_e$.

\subsection{The FP--structure connection}

The FP is often contrasted with the so-called
``virial expectation'', but there is in fact no \textit{a priori}
reason to expect that either the stellar mass-to-light ratio or
the central dark-matter fraction should be constant with
galaxy mass.  Nevertheless, the existence of the FP of elliptical galaxies
implies a regularity in their formation and evolution
history.  Our result explains this regularity in terms
of a universal mass-dynamical structure which is the
end state of massive elliptical evolution regardless
of mass, together with a systematic trend in
total (luminous plus dark) mass-to-light ratio with galaxy mass.
As discussed in \S\ref{mml}, there is no apparent trend of
S\'{e}rsic index $n$ with luminosity within the SLACS sample.
Despite this fact, the SLACS lenses define a clear FP relationship;
the explanation of the SLACS FP must therefore lie with
other factors.  We reiterate here, though, that the SLACS lens sample
is confined to the high-mass end of the elliptical galaxy
population.  Thus we cannot rule out the importance of the
S\'{e}rsic index to the FP of lower-mass early-type galaxies.
An important implication of our result is that the
``dynamical mass'' can be used as a suitable proxy
for the true mass inside 1--10\,kpc, with a conversion factor that is
independent of galaxy mass or size.
A similar result has been reported
by \citet{capp_sauron} for their dynamical
analysis of a mostly lower-mass galaxy sample.

In the strictest sense, the fact that we find no mass-dependent trend
in the ratio of dynamical mass to true mass is suggestive
of a universal mass-dynamical structure, but not fully conclusive.
In principle, a combination of
various other dynamical details---perhaps trends in the
anisotropy profile of the stellar orbits, perhaps trends
in the extent of dynamical relaxation---could conspire
to give an unchanging dimensionless
structure constant with mass.  We also note that
the mass--redshift degeneracy in the SLACS sample
could in principle also be masking a mass-dependent
trend if significant dynamical evolution
occurs with redshift (see, e.g., \citealt{vdm_vd_07a, vdm_vd_07b}),
and the more massive lens galaxies evolve to have the same
dynamical structure as the less massive lens galaxies
at lower redshift.

\subsection{Stellar mass or dark mass?}

Gravitational lensing in both strong and weak
forms measures \textit{total mass}: stars and dark matter
together.\footnote{Gravitational microlensing, in contrast,
can distinguish between mass distributed among point-like
stellar-mass objects and mass distributed in a smooth,
continuous component.  See, e.g., \citet{dw1987, dw1988};
\citet{webster_91}; \citet*{sws94, wms95};
\citet{lewis_irwin_95, lewis_irwin_96, schech_wamb_02}.}
A similar statement holds for mass constraints based upon
dynamical measurements (which we note here even though our use of
stellar dynamics in this paper is limited to empiricism and
dimensional analysis): although the dynamical tracers are
distributed with the optical luminosity,
their orbits are determined by the potential of the total (luminous
plus dark) matter distribution.
The results presented here strongly suggest that the
total mass-density structure of elliptical galaxies
is universal---i.e., not a function of the other observables---at least
over the range of galaxy masses covered by the SLACS sample.
But what is the breakdown of this mass into stellar
and dark-matter components?
Fundamentally, this question cannot be answered
without the imposition of prior conditions on either
the form of the dark-matter
density profile or the mass-to-light ratio of the stellar component.
To express the ambiguity
mathematically, for any stellar and dark-matter density profiles 
$\rho_{\star}$ and $\rho_{\mathrm{DM}}$ that satisfy all observational
lensing and dynamical constraints, we may transform according to
\begin{eqnarray}
\label{dmstars}
\rho_{\star} &\longrightarrow& \rho_{\star}^{\prime} = \beta \rho_{\star} \\
\nonumber
\rho_{\mathrm{DM}} &\longrightarrow& \rho_{\mathrm{DM}}^{\prime}
= \rho_{\mathrm{DM}} + (1-\beta) \rho_{\star}~,
\end{eqnarray}
subject only to the requirement that $\rho_{\star}^{\prime} > 0$
and $\rho_{\mathrm{DM}}^{\prime} > 0$ everywhere,
without altering either the observable quantities or the internal
dynamical consistency of the system.
Previous papers in the SLACS series (Paper~II, Paper~III, Paper~IV) have approached
this problem from both the dark-matter profile and stellar-population
angles.  With forthcoming multi-band photometric data, we will
be able to estimate stellar masses through more detailed stellar-population
modeling, which we may in turn relate to lensing and dynamical
masses to establish overall trends within the population.
\citep[see, e.g.,][]{gallazzi_2006, bundy_2007, grillo_08}.

Qualitatively, our result is in good agreement with the recent
theoretical work of \citet*{bmq_05} and \citet{robertson_06}.  Based on the
analysis of numerical simulations, these authors argue that the
tilt of the FP is the result of a systematic trend in the central
dark-matter fraction that is established and
preserved though the assembly of successively
more massive spheroids via mergers.  \citet{robertson_06}
argue further that this trend is originally established through
the importance of dissipative gas processes in
disk galaxy formation and early merging.
Similar conclusions were
reached observationally by \citet{padmanabhan_04} with the aid
of stellar-population modeling and theoretically-motivated dark-matter
halo models, and by \citet{capp_sauron} through the combination
of spatially resolved dynamical modeling with stellar population modeling.

\subsection{Final thoughts}

In conclusion, our results offer the following physical picture for
the tilt and tightness of the FP for massive early-type galaxies
(see also the discussion of Paper~III).
The luminosity profiles are, on average, well described
by a de Vaucouleurs model luminosity profile.  This profile is embedded
in an on-average scale-free isothermal ($\rho \propto r^{-2}$)
\textit{total} mass density profile, with the distribution of the
dark matter dictated by a bulge-halo ``conspiracy'' to establish
the isothermal total profile.  The
conspiracy is not far-fetched if the isothermal profile is
in fact a dynamical attractor for the evolution of centrally
condensed self-gravitating particle systems,
since stars and cold dark matter together constitute
a single collisionless component
from the point of view of gravitational dynamics \citep{loeb_peebles}.
The galaxies thus form a two-parameter family, indexed by
the effective radius of the stellar distribution (the only
observable scale length) and the central velocity dispersion
of the system (the only parameter of the mass model).
Luminosity (and hence mass-to-light ratio)
then varies systematically
with the scale size of the stellar distribution in such
a way as to give a gradual increase in the total
mass-to-light ratio with increasing total mass within
the stellar effective radius.

Observationally, this picture remains ambiguous as to the systematic
breakdown of mass into stars and dark matter along the
FP, a deficiency that can be partly addressed through
analysis of forthcoming multi-band \textit{HST} imaging
(ACS, WFPC2 and NICMOS)
of the SLACS lens sample, which will give better
insight into the role of stellar-population effects in establishing
the scaling relations examined here.
In addition, spatially
resolved kinematic data are being obtained for the SLACS lenses
\citep{czoske_2321} to gain the tightest possible constraints on the
possible luminous/dark decompositions
\citep[e.g.,][]{barnabe_koopmans}.  This ongoing analysis
will lead to a clearer and more robust picture
of the empirical physical relations described in this work.

\acknowledgments

ASB, TT, LVEK, RG, and LAM acknowledge the support
and hospitality of the Kavli
Institute for Theoretical Physics
at UCSB, where the early stages of
this work were completed.
ASB thanks G. Dobler, G. Novak, and S. Faber for valuable discussion
related to this work.  TT acknowledges support
from the NSF through CAREER award NSF-0642621 and from the Sloan Foundation through a
Sloan Research Fellowship.  He is also supported by a Packard fellowship.
LVEK is supported
in part through an NWO-VIDI program subsidy (project number 639.042.505).
He also acknowledges the continuing support by the European Community's
Sixth Framework Marie Curie Research Training Network Programme,
Contract No. MRTN-CT-2004-505183 (``ANGLES'')\@.
The work of LAM was carried out at Jet Propulsion Laboratory, California
Institute of Technology under a contract with NASA\@.

Support for \textsl{HST} programs \#10174, \#10494,
\#10587, \#10798, and \#10886 was provided by NASA through a grant from
the Space Telescope Science Institute,
which is operated by the Association of Universities for
Research in Astronomy, Inc., under NASA contract NAS 5-26555.
Please see \textsl{HST} data acknowledgment on title page.

This work has made extensive use of the Sloan Digital Sky Survey database.  Funding for the SDSS and SDSS-II has been provided by the Alfred P. Sloan Foundation, the Participating Institutions, the National Science Foundation, the U.S. Department of Energy, the National Aeronautics and Space Administration, the Japanese Monbukagakusho, the Max Planck Society, and the Higher Education Funding Council for England. The SDSS Web Site is \texttt{http://www.sdss.org/}.

The SDSS is managed by the Astrophysical Research Consortium for the Participating Institutions. The Participating Institutions are the American Museum of Natural History, Astrophysical Institute Potsdam, University of Basel, University of Cambridge, Case Western Reserve University, University of Chicago, Drexel University, Fermilab, the Institute for Advanced Study, the Japan Participation Group, Johns Hopkins University, the Joint Institute for Nuclear Astrophysics, the Kavli Institute for Particle Astrophysics and Cosmology, the Korean Scientist Group, the Chinese Academy of Sciences (LAMOST), Los Alamos National Laboratory, the Max-Planck-Institute for Astronomy (MPIA), the Max-Planck-Institute for Astrophysics (MPA), New Mexico State University, Ohio State University, University of Pittsburgh, University of Portsmouth, Princeton University, the United States Naval Observatory, and the University of Washington.

\end{document}